\documentclass[11pt]{llncs}

\usepackage{makeidx}
\usepackage{amsmath}
\usepackage{graphics}
\usepackage{graphicx}
\usepackage{verbatim}
\usepackage{alltt}
\usepackage{amssymb}
\usepackage{ebnf}
\usepackage{mathpartir}
\usepackage{stmaryrd}

\newcommand{\eval}[1]{{$\llbracket$#1$\rrbracket$}$_E$}

\newif\ifpaper

\paperfalse % set paperfalse for the long version (TR)

\begin{document}
\pagestyle{empty}

\title{Pleasantly Consuming Linked Data\\ with RDF Data Descriptions}
 \author{Michael Schmidt\inst{1}\thanks{\textit{Current affiliation:} imc information multimedia communication AG, Scheer Tower, Uni-Campus Nord, 66123 Saarbr\"{u}cken, Germany.} and Georg Lausen\inst{2}}
 \institute{
 fluid Operations AG\\
 Altrottstra{\ss}e 31, 69190 Walldorf, Germany\\
 michael.schmidt@fluidops.com\\[0.1cm]
\and
  University of Freiburg, Institute for Computer Science \\
  Georges-K\"ohler-Allee, 79110 Freiburg, Germany  \\
 lausen@informatik.uni-freiburg.de}
 \maketitle

\begin{abstract}
\vspace{-0.25cm}
Although the intention of RDF is to provide an open, minimally constraining way for representing information, there exists an increasing number of applications for which guarantees on the structure and values of an RDF data set become desirable if not essential. What is missing in this respect are mechanisms to tie RDF data to quality guarantees akin to schemata of relational databases, or DTDs in XML, in particular when translating legacy data coming with a rich set of integrity constraints -- like keys or cardinality restrictions -- into RDF. Addressing this shortcoming, we present the RDF Data Description language (RDD), which makes it possible to specify instance-level data constraints over RDF. Making such constraints explicit does not only help in asserting and maintaining data quality, but also opens up new optimization opportunities for query engines and, most importantly,  makes query formulation a lot easier for users and system developers. We present design goals, syntax, and a formal, First-order logics based semantics of RDDs and discuss the impact on
consuming Linked Data.
\vspace{-0.2cm}
\end{abstract}

%%%%%%%%%%%%%%%%%%%%%%%%%%%%%%%%%%%%%%%%%%%%%%%%%%%%%%%%%%%%%%%%%%%%%%%%%%%%
\section{Introduction}
\label{sec:intro}
\vspace{-.2cm}

Since the early days of relational databases, constraints have been considered essential to specify the intended states of the data sets representing the information of certain applications~\cite{Codd70,Stonebraker75}. In recent years the number of applications that are based on large scale distributed data available on the Internet has been constantly increasing. Many of them are based on RDF~\cite{RDF_W3C} and the question arises, whether relational database like constraints can be considered essential for such applications, as well. RDF data often comes together with RDFS \cite{RDF_W3C} or even OWL \cite{OWL_W3C}, and it is well-known that these languages are not intended to cover relational constraints~\cite{MotikHS07}. Further, designed as rule languages, they do not offer mechanisms to express constraints explicitly over the instance data~\cite{MotikHS09,SequedaAM12,LausenMS08}.

As an example, consider the work on mapping relational databases to RDF from
the W3C's Direct Mapping and R2RML initiatives~\cite{RDB_W3C,R2RML_W3C}.
While they may exploit relational integrity constraints to increase the mapping quality, these
constraints are at most \textit{implicit} in the resulting RDF database: for a data consumer, who may not be
aware of the underlying mappings, no \textit{explicit} guarantees about properties and structure of the data
are available.  For instance, in the Direct Mapping approach~\cite{RDB_W3C} primary keys are exploited to
generate unique IRIs for objects using the key column names and values; yet, there is no constraint in RDF describing
that the properties derived from the key columns are single-valued and identify the resulting objects.
 In fact, designed as rule languages neither RDF(S) nor OWL allow to express constraints~\cite{MotikHS07}. 
Although their built-in semantics may imply certain constraints
(such as type inheritance at instance-level for \textit{rdfs:subClass} relationships), constraints are
only \textit{implicit} and, moreover, may not hold when the data is published under ground semantics --
a common scenario in the Linked Data context.

{\bf Contributions.} After motivating the need to enable end users in writing precise SPARQL queries in Section~\ref{sec:example},
we present the {\em RDF} {\em Data} {\em Description} language, RDD, to define constraints over RDF,
akin to DTDs for XML. We then discuss design decisions and related work in Section~\ref{sec:designgoals},
identifying the need for RDDs to be both \textit{user-readable} and \textit{machine-processable}.
Next, we elaborate on the conflict between the Open World Assumption underlying RDF(S) and the
requirements of a hard constraint language, concluding that RDDs shall support a \textit{pay-as-you-go paradigm}
in constraining RDF(S). Section~\ref{sec:syntax} formalizes RDDs by means of  a \textit{user-friendly syntax} that
captures a broad range of constraints including keys, cardinalities, subclass, and subproperty restrictions.
Section~\ref{sec:semantics} presents a \textit{First-order Logics semantics}, making it easy to implement
RDD checkers and clearing the way for optimizations. Finally, in Section~\ref{sec:discussion} we discuss
\textit{directions of future research}, including the implementation, coverage, extensibility, and relationship
to standards like VoID~\cite{VoID}.

\section{RDD by Example}
\label{sec:example}
\vspace{-.2cm}

As a motivating example, assume a developer wants to write a SPARQL query
that extracts information about persons in an RDF document, described by properties {\tt rdfs:label} (denoting the name),
{\tt foaf:age}, and {\tt foaf:mbox} (mail address) -- where  every person shall be represented
by exactly one row of the result table.\footnote{In fact, the problem is a slightly modified example one of the authors
recently encountered in the context of an industrial project.} While this sounds like a fairly trivial task
(in SQL, with a reasonable schema, this could probably be expressed by
a simple query like \verb!SELECT id, name, age, email FROM Person!), with the unconstrained RDF model this may become
quite tricky, even if the schema (i.e., FOAF and RDF(S) vocabulary) is well known to the developer: without further knowledge
about the \textit{instance data}, the developer cannot be sure which predicates are present at all, and which of
them may be multi-valued. Making guesses that {\tt rdfs:label} and {\tt foaf:age} are single-valued,
the developer may finally come up with the following query:

{\scriptsize
\begin{verbatim}
  SELECT ?person ?name ?age (GROUP_CONCAT(?mail; separator=", ")  AS ?mail)
  WHERE {  ?person rdf:type foaf:Person .
            OPTIONAL { ?person rdfs:label ?name }
            OPTIONAL { ?person foaf:age ?age }
            OPTIONAL { ?person foaf:mbox ?mail }   } GROUP BY ?person ?name ?age
\end{verbatim}
}

The \textsc{OPTIONAL} clauses ensure that persons with incomplete information
are included in the result; to group persons with multiple email addresses, the
developer used \verb!GROUP BY! combined with \verb!GROUP_CONCAT!
in the \verb!SELECT! clause, thus concatenating all email addresses of a single person.
The crucial point here is that even this simple task leads to a quite complex query
covering the ``worst case scenario'' anticipated by the developer, requiring the use of advanced SPARQL 1.1
constructs (which, as a matter of fact, are hard to optimize by query engines). And even this carefully
designed query leads to multiple result rows for the same person in the presence of multiple labels (e.g.,
with different language tags).

What is needed to ease SPARQL query development is a data description that describes the structural constraints of
the instance data \textit{beyond} the schema information contained in the underlying RDF(S) specification
and ontologies, which the developer can consult when writing queries. The RDD language advocated in
this paper was designed with exactly this goal in mind. RDD would allow the data publisher to express the
instance data constraints by means of a well-defined, human readable language.

Figure~\ref{fig:example}
depicts an example RDD that, when tied to a specific RDF database, helps the developer in 
understanding the constraints that hold on instance level. With respect to the concept \textit{foaf:Person},
the first part of the RDD in Figure~\ref{fig:example} (left) specifies a set of constraints that are known to
hold for every instance of the class. Summarizing the relevant part of the RDD, it tells the developer
that the property \textit{rdfs:label} serves as a key for persons, every person has exactly one
\textit{foaf:mbox} property (keyword \verb!TOTAL!), and every person has at most one \textit{foaf:age}
(keyword \verb!PARTIAL!). Further, all theses three properties point to literals, the latter being of
type \textit{xsd:integer} -- this may be useful information when writing
e.g.~aggregation queries over the age, or when post-formatting the results.

\begin{figure}[ht!]
\hspace{-1.3cm}
\begin{minipage}{18cm}
\begin{tabular}{l|ll}
\begin{minipage}{7cm}
{\scriptsize
\begin{verbatim}
PREFIX ex: <http://www.example.com#>
...

CWA CLASSES {
 OWA CLASS foaf:Person SUBCLASS ex:Student {
   KEY rdfs:label : LITERAL;
   TOTAL foaf:mbox : LITERAL;
   PARTIAL foaf:age : LITERAL(xsd:integer);
   RANGE(foaf:Person) foaf:knows : IRI; }
\end{verbatim}
}
\end{minipage}
&
\ \ \ 
&
\begin{minipage}{7.2cm}
{\scriptsize
\begin{verbatim}
 OWA CLASS ex:Student {
   TOTAL ex:matricNr : LITERAL(xsd:integer);
   MIN(1), RANGE(ex:Course) ex:course : RESOURCE;
   PATH(ex:course/ex:givenBy), RANGE(foaf:Person) 
     ex:taughtBy : IRI; }
}

OWA PROPERTIES {
  TOTAL rdfs:label;
  foaf:knows SUBPROPERTY ex:taughtBy; }
\end{verbatim}
}
\end{minipage}
\end{tabular}
\end{minipage}
\caption{Example RDF Data Description}
\vspace{-0.2cm}
\label{fig:example}
\end{figure}

With the RDD specification at
hand -- which can be understood in few seconds -- the developer can considerably simplify the query:

{\scriptsize
\begin{verbatim}
  SELECT ?person ?name ?age ?mail .
  WHERE { ?person rdf:type foaf:Person ; rdfs:label ?name ; foaf:mbox ?mail .
          OPTIONAL { ?person foaf:age ?age } }
\end{verbatim}
}

Even if the developer is not aware of the RDD and comes up with a query that
uses, e.g., redundant \textsc{OPTIONAL} blocks, the RDD 
may still be used by the optimizer to simplify the query and speed up evaluation.

\section{Design Decisions and Related Work}
\label{sec:designgoals}
\vspace{-0.18cm}

{\bf Philosophy.} RDDs specify constraints that hold in an RDF data set.  However, not to loose RDF's minimally
constraining way for representing information -- where one may interlink and extend data sets by adding
new information -- they shall not require the structure of RDF to be defined \textit{completely},
but give a pragmatic answer to these two conflicting design goals in that they
adhere to RDF's Open World character following a \textit{pay-as-you-go paradigm},
%adhere to RDF's Open World character following a \textit{pay-as-you-go paradigm},
which allows users to impose constraints only on a subset of classes, or to constrain classes and properties only partially.

\smallskip
{\bf Designed for Humans.} 
To make it easy for humans to understand, write, and use RDDs as a guide when writing queries,
RDDs shall come with a \textit{user-understandable} syntax. To this end, we use an object-oriented
approach closely aligned to the RDF(S) data model, reusing concepts like classes, properties, and subclass/subproperty
relationships. An RDF serialization of RDDs is out of the scope of this paper (cf.~Section~\ref{sec:discussion}).

\smallskip
{\bf Scope.} The importance of constraints for RDF(S) has recently been emphasized in the
context of REST-based enterprise applications~\cite{OSLC}. With the goal to provide a
\textit{machine-readable} language, \textit{OSLC Resource Shape} defines an RDF vocabulary
to encode qualified property constraints (such as cardinality, range, or value restrictions).
While no formal semantics is given, the authors propose an implementation via SPARQL \textsc{ASK}
queries. RDDs, in contrast, are designed for humans, come with a formal semantics, and 
go far beyond what can be expressed with OSLC (e.g., expressing completeness guarantees and
unqualified property constraints). 

Enabling the targeted restriction of RDF(S) constructs,
RDDs provide built-in constructs to express constraints over classes, subclasses, and
properties such as domain, range, or cardinality restrictions. 
In the light of the Direct Mapping and R2RML standards~\cite{RDB_W3C,R2RML_W3C},
RDDs shall also cover constraints from the relational databases domain, in order
to carry over integrity information when translating relational data.

RDF data is often equipped with RDFS or OWL axioms and may be interpreted in different
entailment regimes. Dedicated studies of constraints in the context of OWL have been presented
in~\cite{MotikHS09,DBLP:conf/semweb/Tao10}. Adhering to the different semantics under which RDF
can be published, RDDs should be independent from the entailment regime. Our approach is similar to
SPARQL~\cite{SPARQL_Entailment}, which also supports different entailment regimes 
(and is defined independently): if, e.g., RDF is published under ground semantics,
an associated RDD spec would specify the constraints that hold on the bare instance data;
if, e.g., RDFS inferencing is
turned on, an RDD specification would take inferred facts into account -- in both cases, an end user
can transparently rely on the RDD spec when accessing the data.

\smallskip
{\bf Formal Semantics.}
While a SPARQL-based semantics may seem like a natural choice (cf.~\cite{OSLC,LausenMS08}), we argue that is desirable to
choose a semantics that can easily be mapped to existing work on integrity constraints from the relational database community,
e.g.~to carry over Semantic Query Optimization techniques (e.g., the seminal work~\cite{DBLP:journals/tods/MaierMS79}).
We therefore decided for a First-order Logics (FOL) based semantics, representing constraints
as First-order sentences known as tuple-generating and equality-generating dependencies~\cite{beeri1984formal},
which are well understood from previous investigations (e.g.~\cite{DBLP:journals/tods/MaierMS79}). A possible
implementation of our FOL based semantics by means of SPARQL will be discussed later in Section~\ref{sec:discussion}.

\section{RDD Syntax and Model}
\vspace{-0.3cm}
\label{sec:syntax}
Figure~\ref{fig:example} provides an example RDD. The definition for class \textit{foaf:Person}
contains the constraints for predicates \textit{rdfs:label}, \textit{foaf:email}, and \textit{foaf:age} discussed
in Sec.~\ref{sec:example}, plus a constraint expressing that predicate \textit{foaf:knows}, when used
for an instances of type \textit{foaf:Person}, points to instances of type \textit{foaf:Person},
which are always IRIs (i.e., not blank nodes). In the spirit of RDF, this does not enforce 
referred objects to be \textit{exclusively} typed as \textit{foaf:Person}, but only that one edge
typing the object as \textit{foaf:Person} is present. 
The \verb!OWA! keyword (short for \textit{Open
World Assumption}) in front of the \verb!CLASS! definition allows persons to have further properties
not listed in the class specification; its counterpart, keyword \verb!CWA!, would enforce that the class
is \textit{closed} in the sense that class instances are \textit{completely} described by the properties
occurring in the \verb!CLASS! section. Further,  \textit{foaf:Person} has a subclass \textit{ex:Student}
(keyword \verb!SUBCLASS!). With RDDs focusing on instance-level constraints, this does
\textbf{not} enforce a triple (\textit{ex:Student}, \textit{rdfs:subClassOf}, \textit{foaf:Person}) in the data,
but guarantees that every \textit{ex:Student} satisfies the same constraints as \textit{foaf:Person}s.

Similar in spirit, the \verb!CLASS! definition for \textit{ex:Student} guarantees that
(i)~\textit{ex:matricNr} is a \verb!TOTAL! property, i.e.~every student has exactly one matriculation number,
(ii)~\textit{ex:course} occurs at least once and has \verb!RANGE! \textit{ex:Course},
and (iii)~the \verb!PATH! and \verb!RANGE! constraints defined for property \textit{ex:taughtBy} asserts
that, for every \textit{ex:Student}, there is a property path along the edges \textit{ex:course} followed
by \textit{ex:givenBy} pointing to the same value of type \textit{foaf:Person} as property \textit{ex:taughtBy}. Path
constraints are special kinds of inclusion constraint, similar in spirit to foreign keys.

The surrounding \verb!CLASSES! section is defined as \verb!CWA!, guaranteeing completeness in the sense
that the data contains only instances of the two classes \textit{foaf:Person} and \textit{ex:Student}.
Subsequently, the RDD contains a \verb!PROPERTIES! section constraining properties in
an \textit{unqualified}, \textit{global} way.
The first entry, \verb!TOTAL rdfs:label!, ensures that \textit{every Resource} 
has exactly one label -- a useful information for query authors. 
The \verb!SUBPROPERTY! spec, targeting the instance level again,
ensures that, for every triple (X,\textit{ex:taughtBy},Y) in the data, there is
an implied triple (X,\textit{foaf:knows},Y). Finally, the \verb!OWA! keyword of
the \verb!PROPERTIES! section expresses that there may be other properties than
those listed in the section.

%%%%%%%% SWITCH REFERRING EITHER TO APPENDIX OR TECHNICAL REPORT
\ifpaper % SHORT PAPER
Rather than presenting the concrete syntax (see the Technical Report~\cite{sl2013} for this level of detail),
\else % LONG PAPER (TR)
A formal grammar for the concrete syntax is presented in Appendix~\ref{app:grammar},
\fi
Figure~\ref{fig:abstractsyntax} visualizes the structure and concepts of the RDD language
in a UML-style notation. Boxes denote concepts, arrowed lines sub-concept relationships, and the
second line type indicates that concept $A$ uses $B$. At top-level, RDDs consist of a \textsc{ClassConstraintSec}
and a \textsc{PropConstraintSec}, which contain lists of \textsc{ClassConstraint}s and \textsc{PropConstraint}s,
respectively, plus a boolean flag indicating whether the sections should be interpreted under Open or Closed World Assumption
(i.e., whether the classes and properties in the sections describe the RDF data completely or nor)
-- the \verb!OWA! keyword enables RDDs to be written in a pay-as-you-go fashion, where known constraints
are specified, while unknown parts are left unspecified.

Central to the RDD concept is the notion of \textsc{PropConstraint}s, an abstract concept that is further subclassed
into specific subclasses:

\begin{figure}[ht]
\vspace{-0.4cm}
\hspace*{-1.2cm}
\includegraphics[width=14.5cm]{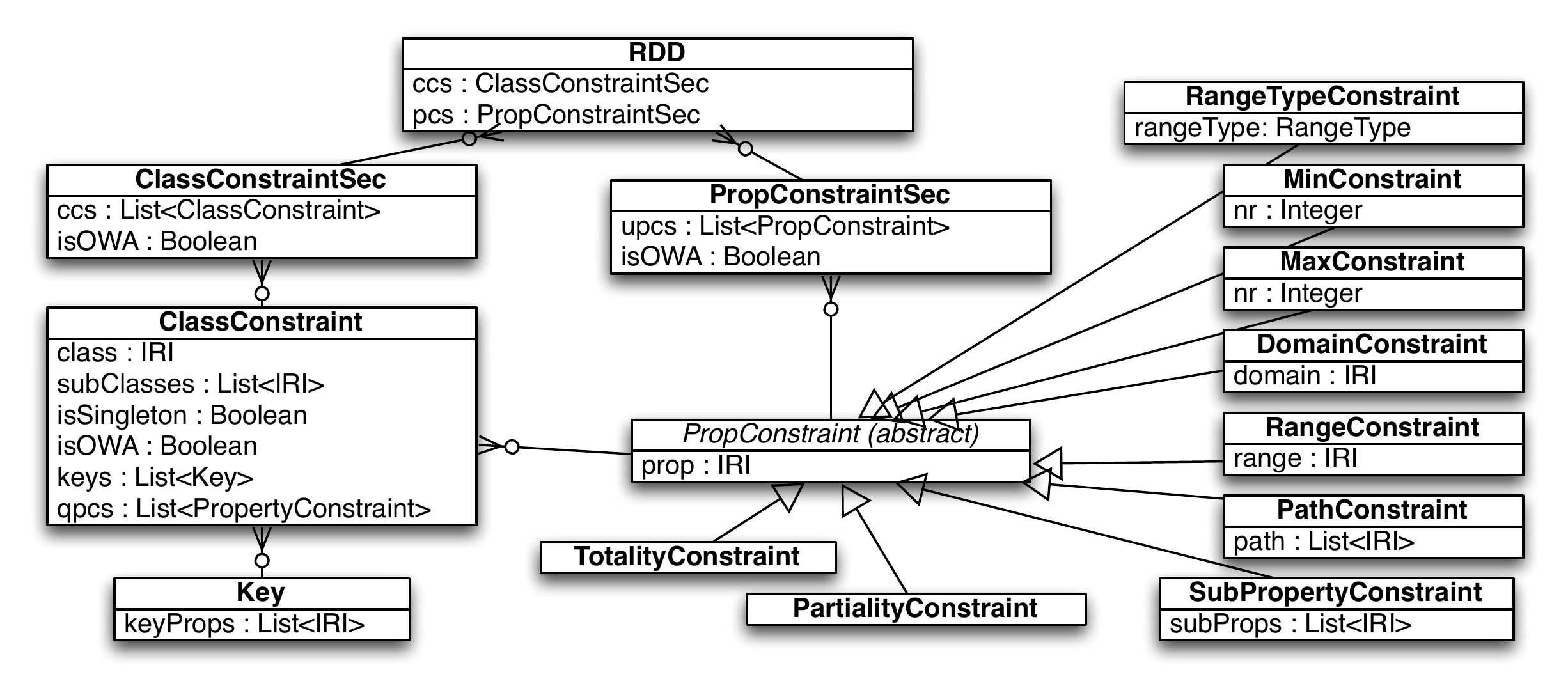}
\vspace{-1cm}
\caption{Structural Overview of the RDD Language}
\vspace{-0.2cm}
\label{fig:abstractsyntax}
\end{figure}

\vspace{-0.1cm}
\begin{itemize}
\item A \textsc{RangeTypeConstraint} indicates that the property \textit{prop} points to either a URI, BlankNode, Resource, or a (possibly typed) Literal.
\item A \textsc{Min/MaxConstraint} indicates that the property \textit{prop} occurs at least or at most \textit{nr} times, respectively.
\item A \textsc{Domain/RangeConstraint} indicates a guaranteed \textit{domain} or \textit{range} for subject and objects associated with \textit{prop}, respectively.
\item A \textsc{PathConstraint} indicates that the value of \textit{prop} can as well be reached by following a given \textit{path} of properties.
\item A \textsc{SubPropertyConstraint}  indicates that for every triple using property \textit{subProp},
there is also an identical triple using property \textit{prop}.
\item \textsc{Partiality/TotalityConstraint}s express that \textit{prop} occurs at most or exactly one time, respectively.
\end{itemize}
\vspace{-0.1cm}

\textsc{PropConstraint}s are used in two different contexts: (1) The \textsc{PropConstraintSec}
(cf.~keyword \verb!PROPERTIES!) contains a list \textit{upcs} of \textsc{PropConstraint}s, implementing
\textit{unqualified}, global characteristics of properties. For instance, the \textsc{TotalityConstraint}
in Fig.~\ref{fig:example} (keyword \verb!TOTAL!) for $prop:=$\textit{rdfs:label} asserts that \textit{every resource}
has exactly one label. (2) Variable \textit{qpcs}  inside \textsc{ClassConstraint}s 
represents \textit{qualified}, class-specific \textsc{PropConstraint}s, e.g.~the
\textsc{MinConstraint} (keyword \verb!MIN!) for $prop:=$\textit{ex:course} and $nr:=1$
in Fig.~\ref{fig:example} in the class section of \textit{ex:Student} ensures that
\textit{every instance of} \textit{ex:Student} visits at least one course. 

In addition to \textit{qpcs}, a \textsc{ClassConstraint}  contains
(i) a list of subclasses (keyword \verb!SUBCLASS!), enforcing that instances of the subclasses inherit
 inner constraints of the superclass, (ii) a boolean flag \textit{isSingleton}, enforcing that exactly one
instance of the class exists, (iii)~the \textit{isOWA} flag, and (iv)~a list of keys. We sketch their semantics in the next section.

\section{RDD Semantics}
\label{sec:semantics}
\vspace{-.2cm}

The semantics, denoted by \eval{r}, decomposes an RDD $r$ into constraints that can be
checked individually and independently. It uses an environment~$E$ capturing
\verb!SUBCLASS! and \verb!SUBPROPERTY! relations specified in the RDD. The result is a set of
in First-order Logics (FOL) constraints over relation $T_D(s,p,o)$ representing the RDF
triples in  RDF data set $D$.  With~\eval{.} at hand, we define the notion of \textit{consistency} as follows.

\begin{definition} \em
Let $D$ be an RDF data set and $r$ be an RDD specification. Further
let $cs$ := \eval{r} be the set of first-order logic constraints defined by $r$.
Data set $D$ is \textit{consistent} w.r.t.~$r$ if and only if for all
constraints~$c \in cs$ it holds that $c$ is valid in $T_D$, i.e.~$T_D \models c$.
\end{definition}

%$E.C: IRI \mapsto List\<IRI>$ we denote the function that maps a class name to its subclasses as specified
%by the \verb!SUBCLASS! keywords of the RDD; $E.A: IRI \mapsto List\<IRI>$ denotes a function that,
%given a class $C$ as input, returns all the properties listed in the class section.
%For the example from Figure~\ref{fig:example} we have $E.C(\textit{foaf:Person}) = \{ \textit{ex:Student}\ \}$
%and for all other $x \in IRI: E.C(x) = \emptyset$.  

The evaluation function \eval{.} is defined by about $40$ rules along the structure
of RDDs. At top-level, an RDD is decomposed into its \textsc{ClassConstraintSec} and
\textsc{PropConstraintSec} (i.e., the members of class \textsc{RDD}, cf.~Fig.~\ref{fig:abstractsyntax}),
which are then further decomposed by dedicated rules. At the core are inference rules mapping the
individual constraints -- e.g., the \textsc{PropConstraint} subclasses -- into FOL. 
%%%%%%%% SWITCH REFERRING EITHER TO APPENDIX OR TECHNICAL REPORT
\ifpaper % SHORT PAPER
We sketch the idea of the evaluation and refer to
the TR~\cite{sl2013} for a complete listing of the rules.
\else % LONG PAPER (TR)
In this section, we sketch the basic idea of the system. The interested reader will find 
a complete listing and discussion of the inference system in Appendix~\ref{app:semantics}.
\fi

Let us exemplarily discuss the inference rule for \textsc{ClassConstraint}s, which derives constraints
for its key and property constraints, as well as its global ``configuration'',
namely, the subclass hierarchy, whether it is defined as singleton,
and whether it is defined under Open or Closed World Assumption. According to Figure~\ref{fig:abstractsyntax},
the \textsc{ClassConstraint} is represented by a structure
(\textit{class},\textit{subClasses},\textit{isSingleton},\textit{keys},\textit{qpcs},\textit{isOWA}).

{\small
\[
\inferrule{cs_{singleton} := $\eval{(\textit{class}, \textit{isSingleton}) : Singleton}$ \\
               cs_{sc} := \bigcup_{c_{sc} \in \textit{subClasses}} $\eval{($c_{sc}$,\textbf{false},$E.C$($c_{sc}$),\textit{keys},\textit{cpcs},\textbf{true}) : ClassConstraint}$ \\
               cs_{key} := \bigcup_{key \in keys} $\eval{(\textit{class},\textit{key}) : ClassKey}$ \\ 
               cs_{qpcs} := \bigcup_{cpc \in cpcs} $\eval{(\textit{class},\textit{cpc}) : ClassPropConstraint}$ \\
               cs_{wa} := $\eval{(\textit{class\textit}, $E.A$(\textit{class}), \textit{isOWA}) : OWA$_P$}$}
          {\llbracket$(\textit{class} : IRI, \textit{subClasses} : List\<IRI>, \textit{isSingleton} : Boolean, \textit{keys} : List\<Key>,$ \\
             $\textit{qpcs} : List\<PropConstraint>, \textit{isOWA} : Boolean) : ClassConstraint$\rrbracket \vdash \\
             cs_{singleton} \cup cs_{sc} \cup cs_{key} \cup cs_{qpcs} \cup cs_{wa}}
\]
}

Starting with the conclusion, the result of evaluating the \textsc{ClassConstraint} is the union of
the constraint sets $cs_{singleton}$, $cs_{sc}$, $\dots$; the premise of the rule describes how these constraint
sets are calculated. The constraint set $cs_{singleton}$, for instance, is obtained by evaluating a substructure
\textit{Singleton} with \textit{class} IRI and the \textit{isSingleton} flag as argument -- if  \textit{isSingleton}=\textbf{true},
this substructure generates a constraint enforcing that the class has exactly one instance. The scheme for computing
$cs_{wa}$, $cs_{key}$ and the qualified \textsc{PropConstraint}s $cs_{qpcs}$ are analogous. Most interesting is the
computation of $cs_{sc}$, which captures the inheritance of constraints to subclasses. It is obtained by evaluating a
replicated version of the class constraint for every subclass $c_{sc}$. In these replicas, we pass the \textit{keys} and
\textit{cpcs} constraints, and consult environment $E$ to obtain the subclasses of the $c_{sc}$. Note that
we neither inherit the singleton constraint (passing \textit{isSingleton}:=\textbf{false}) nor impose
a \verb!CWA! constraint on the subclass (passing \textit{isOWA}=\textbf{true}); this gives the RDD designer greater flexibility.

To conclude, let us  sketch the evaluation of a qualified \textsc{RangeConstraint}
as one of the constraint-generating rules. The rule below implements such a constraint for
property $p$ with range $R$ for class $C$: the resulting formula enforces that for
every instance $s$ of class $C$, every value $o$ referenced by $p$ is of type $R$. Note the difference
toward RDFS: while predicate \textit{rdfs:range} is similar by idea, it does not enforce the presence of such a triple,
but sets up a rule that generates/completes the data -- the constraint \textit{guarantees} 
the presence of such a triple in the database.

{\small
\[
\inferrule{ }
          {$\eval{($C$ : IRI, ($p$ : IRI, $R$ : IRI) : RangeConstraint)}$\ \vdash \\ \{ \forall s, o (T_D(s,\textbf{rdf:type},C) \land T_D(s,p,o) \rightarrow T_D(o,\textbf{rdf:type},R)) \}}
\]
}
\vspace{-.3cm}

\section{Discussion and Future Research Direction}
\label{sec:discussion}
\vspace{-.2cm}

RDDs constitute a powerful mechanism to describe instance-level constraints and help 
both humans -- when writing SPARQL queries -- and engines -- which may exploit RDDs
with its concise semantics to assert data consistency and optimize queries (cf.~\cite{LausenMS08}).
Our approach opens up a new research field, which we will shortly discuss in the following.

\smallskip
\smallskip
{\bf Implementation.}
Given the First-order semantics, it is straightforward to build a constraint checker
using existing FOL engines. In the context of SPARQL engines, though, it may be favorable
to implement checkers by means of SPARQL \textsc{Ask} queries. The close connection
between SPARQL and logics-based formalisms has been pointed out in several 
works~\cite{DBLP:conf/www/Polleres07,DBLP:conf/icdt/Schmidt0L10}. In~\cite{LausenMS08} it was
shown how to encode constraints such as keys and
cardinalities in SPARQL. Further, it was proven in~\cite{DBLP:conf/icdt/Schmidt0L10} by a
constructive proof that every First-order sentence can be expressed in SPARQL. These results make
it easy to map the FOL semantics to SPARQL.
The efficient implementation of constraint checking, though, is a challenging task left for future work.
A simple approach based on a one-by-one execution of \textsc{Ask} queries may have limitations,
in particular when it comes to scenarios with frequent data updates. Here, incremental
constraint checking approaches would be required. However, given that integrity constraints are an
integral part of other data models (such as relational data and DTDs for XML) and that the logical
structure of RDD constraints -- EGDs and TGDs, which are well understood in theory -- is very similar,
we are convinced that efficient implementations are possible.

\smallskip
{\bf Deriving RDDs.}
An interesting topic is the derivation of  RDDs from instance data or -- when the data
has been obtained from relational systems~\cite{RDB_W3C,R2RML_W3C} -- from constraints in
the original data sources.  RDFS or OWL specifications, for instance, that are included in the data,
or query logs may give valuable hints about candidate constraints that hold in the instance data,
allowing to automatically build RDDs that could be refined manually. In this line, it would also be interesting to
study interrelations between different entailment regimes and their implications for RDDs.

\smallskip
{\bf Coverage and Extensibility.}
Since the early years of database research, various classes of constraints have been
investigated (see e.g.~\cite{a1974,DBLP:journals/tods/MaierMS79}).
Based on the design goals from Section~\ref{sec:designgoals}, we selected a reasonable set of
constraints that (i)~may be encountered in typical RDF(S) scenarios and (ii)~may be of benefit when writing
SPARQL queries. There are, of course, constraint types that are currently not supported by RDDs.
One extension would be user-defined constraints in the form of arbitrary SPARQL \textsc{Ask} queries --
they could be easily added through a new \verb!USER! in RDDs,
containing a list of queries including the expected results. While such custom constraints may be
hard to understand by users, they could help to model complex data consistency scenarios.
Other types of constraints that are candidates for extension include functional dependencies~\cite{codd1972further},
EGDs and TGDs with disjunction~\cite{DBLP:conf/pods/MeierSWL10},\footnote{In RDDs, 
some constraints like the \textsc{CWA} restriction implicitly contain disjunction. We may want to
express disjunction in other contexts, e.g. over range restrictions.}
or value restriction (e.g.~expressing that property \textit{foaf:gender} maps to either \textit{male} or \textit{female}).

As a side note, we want to point out that we intentionally did not include foreign keys as proposed
in~\cite{LausenMS08}. Although they play an important role in relational modeling,
they are an artificial construct arising due to the relational structure: when mapping relational database
into RDF(S), foreign keys typically result in range specifications over object properties
(cf.~\cite{SequedaTCM11,RDB_W3C}), which RDDs can easily capture using \textsc{RangeConstraint}s.

\smallskip
{\bf Other applications.} RDDs can be exploited for use cases beyond query formulation, semantic query optimization,
and quality assurance. For instance, the constraints encoded in RDDs may give valuable input
for schema mapping and alignment. As another example, RDDs could be used to derive
precise data input forms, and thus help in producing data.

\smallskip
{\bf RDD and Linked Open Data.}
Constraint checking, at first glance, may look like a local task. However, as RDF resources being
identified by IRIs are globally unique, local checking of constraints for certain constraints
classes may not be sufficient. Assume an inclusion dependency, say a range constraint $r$,
is violated in data set $R_1$, however the violating resource exists and is typed accordingly
in a data set $R_2$. According to the principles of Linked Open Data, $r$ should not be declared to be violated.

%Assume a resource $a$ fulfills all the constraints applicable on $a$ locally in dataset $R_1$, however
%there exist a constraint $C$, which $a$ does not fulfill, if $C$ would be applied on $a$. Now assume a data
%set $R_2$ contains subclass-definitions, whose application on $R_1$ would imply inheritance of $C$ to the
%class $a$ belongs to. As a consequence, data set $R_1$ does not fulfill all the stated constraints. 

We conclude that constraint checking in an LOD environment in general requires us to consider the constraints
of all involved data sets. Even though this does not make constraint checking inapplicable, it may require novel
ways and paradigms to specify and check constraints in the context of an open world, with possibly incomplete
knowledge. We leave a closer investigation of these issues for future work.

\smallskip
{\bf Publishing of RDDs.}
 As a bridge between the publishers and users of RDF data, W3C proposes the VoID vocabulary (\textit{Vocabulary
of Interlinked Datasets})~\cite{VoID,VoID_W3C}. We suggest to develop a canonical RDF representation for RDDs
(coexisting with the user-friendly syntax presented in the paper), with tooling to convert between the two syntaxes.
This would make it quite easy to, e.g., publish RDD descriptions as part of VoID. A detailed study of the relationships toward
VoID and an RDF serialization for the RDD language are interesting topics for future work.

\smallskip
{\bf Standardization.}
 The next steps we plan are the implementation of the RDD language by means of a SPARQL query generator that outputs the \textsc{Ask} queries for checking RDDs, which would make RDDs immediately usable by any SPARQL engine. Further, we are investigating different ways to standardize the proposed language, e.g.~as part of the W3C standardization activities in the semantic technology space.

 \subsubsection{Acknowledgments.}
This work was supported by the German Federal Ministry of Economics and Technology as part
of the project ``\textit{Durchblick}'', grant KF2587503BZ2, and by Deutsche Forschungsgesellschaft as part of the
project ``\textit{CORSOS}'', grant LA 598/7-1. 

The authors want to thank Peter Haase and Michael Meier for fruitful
discussions and the anonymous reviewers for their constructive feedback.

%\section{Related Work}
%\label{sec:related}
%\input{related}

%\section{Conclusion}
%\label{sec:conclusion}
%\input{conclusion}

% % % % % % % % % % % % % % % % % % % % % % % % % % % % % % % % % % % % % % % %

{\small \bibliography{RDD}
\bibliographystyle{IEEEtran}}

\ifpaper
  % no appendix
\else
 \newpage
 \begin{appendix}
 \section{RDD EBNF Grammar}
 \label{app:grammar}
 The start symbol is \<RDD>, terminals are marked in \textbf{bold} font.

\bigskip

{\small
\begin{EBNF}
    \item[RDD]
	\<PrefixDecl>* \<ClassConstraintSec> \<PropConstraintSec>

    \bigskip
    \item[ClassConstraintSec] 
        \textbf{\<WA> \textsc{CLASSES}} \textbf{\textsc{\{}} \<ClassConstraint>* \textbf{\textsc{\}}}
    \item[PropConstraintSec] 
	\textbf{\<WA> \textsc{PROPERTIES}} \textbf{\textsc{\{}} \<PropConstraint>* \textbf{\textsc{\}}}
    \medskip
   \item[ClassConstraint] 
	\<WA> (\textbf{\textsc{SINGLETON}})? \textbf{\textsc{CLASS}} \<IRI>\\
        \quad\quad\quad\quad\quad\quad\quad\quad\quad  (\textbf{\textsc{SUBCLASS}} \<IRIList>)?\\
        \quad\quad\quad\quad\quad\quad\quad\quad\quad \textbf{\textsc{\{}}  (\<Key> | \<PropConstraint>)* \textbf{\textsc{\}}}
    \item[Key]
        \textbf{\textsc{KEY}} \<IRIWithRangeTypeList> \textbf{\textsc{;}}
    \item[PropConstraint]
        \<ConstraintList>? \<IRIWithRangeType> \textbf{\textsc{;}}
    \medskip
    \item[ConstraintList] 
        \<Constraint> (\textbf{\textsc{,}} \<Constraint>)*
    \item[Constraint] 
        \<MinConstraint> | \<MaxConstraint> |\\
        \quad\quad\quad\quad\quad\quad \<DomainConstraint> | \<RangeConstraint> |\\
        \quad\quad\quad\quad\quad\quad \<PartialityConstraint> | \<TotalityConstraint> |\\
        \quad\quad\quad\quad\quad\quad \<PathConstraint> | \<SubPropertyConstraint>
    \medskip
    \item[MinConstraint]
        \textbf{\textsc{MIN(}}\<INTEGER>\textbf{\textsc{)}}
    \item[MaxConstraint]
        \textbf{\textsc{MAX(}}\<INTEGER>\textbf{\textsc{)}}
    \item[DomainConstraint]
       \textbf{\textsc{DOMAIN(}}\<IRI>\textbf{\textsc{)}}
    \item[RangeConstraint]
        \textbf{\textsc{RANGE(}}\<IRI>\textbf{\textsc{)}}
    \item[PathConstraint]
        \textbf{\textsc{PATH(}}\<IRISeq>\textbf{\textsc{)}}
    \item[SubPropertyConstraint]
        \textbf{\textsc{SUBPROPERTY(}}\<IRIList>\textbf{\textsc{)}}
    \item[PartialityConstraint] \textbf{\textsc{PARTIAL}}
    \item[TotalityConstraint] \textbf{\textsc{TOTAL}}
    \item[WA] \textbf{OWA} | \textbf{CWA}

   \bigskip
    \item[IRIList]
        \<IRI> (\textbf{\textsc{,}} \<IRI>)*
    \item[IRISeq]
        \<IRI> (\textbf{\textsc{/}} \<IRI>)*
    \item[IRIWithRangeTypeList]
        \<IRIWithRangeType> \\
        \quad\quad\quad\quad\quad\quad\quad\quad\quad\quad\quad\quad (\textbf{\textsc{,}} \<IRIWithRangeType>)*
    \item[IRIWithRangeType]
        \<IRI> (\textbf{\textsc{:}} \<RangeType>)?
    \item[RangeType]
        \textbf{\textsc{IRI}} | \textbf{\textsc{BNODE}} | \textbf{\textsc{RESOURCE}} | \textbf{\textsc{LITERAL}}(\textbf{\textsc{(}}\<IRI>\textbf{\textsc{)}})?

   \bigskip
    \item[PrefixDecl] as defined in rule~\textit{[6]~PrefixDecl} in \cite{SPARQL_W3C}
    \item[IRI] as defined in rule~\textit{[136]~iri} in \cite{SPARQL_W3C}
    \item[PrefixedName] as defined in rule~\textit{[137] PrefixedName} in \cite{SPARQL_W3C}
    \item[IRIREF] as defined in rule~\textit{[139]~IRIREF} in \cite{SPARQL_W3C}
    \item[INTEGER] as defined in rule~\textit{[146] INTEGER} in \cite{SPARQL_W3C} 
%    \item[AskQuery] as defined in rule~\textit{[12] AskQuery} in \cite{SPARQL_W3C}
\end{EBNF}
}

 \newpage
 \section{RDD Semantics}
 \label{app:semantics}
 \subsection{Prerequisites}
\label{app:prerequisites}
The semantics, denoted by \eval{r : RDD}, decomposes an RDD $r$ into constraints that can be
checked individually and independently. The result is a set of in First-order Logics
constraints over a single relation $T_D(s,p,o)$ representing the set of RDF triples in some RDF
data set $D$. We assume~$D$ to be fixed and shall simply write $T$ instead of $T_D$ in the
following.

In addition to $T$, we assume four unary relations $IRI(x)$, $BNode(x)$, 
$Resource(x)$, and $Literal(x)$ containing all IRIs, blank nodes, resources (i.e., IRIs plus
blank nodes), and literals that appear in any position of any tuple in $D$, respectively.
The semantics will be defined along the concepts of the grammar presented in
Appendix~\ref{app:grammar}. Focusing on the core ideas, we omit the
formalization of technicalities such as the resolving of namespace prefixes or
datatype verification for literals.

To improve readability, we define the following two shortcuts:

\begin{tabbing}
xxx \= \kill
\>$\textit{allDist}(x_1,\dots,x_n) := \bigwedge_{1\leq i<j\leq n}x_i \neq x_j$, and\\
\>$\textit{someEq}(x_1,\dots,x_n) := \bigvee_{1\leq i<j\leq n}x_i = x_j$,
\end{tabbing}

enforcing that the $n$ elements passed as parameters are pairwise
distinct (\textit{allDist}) or not all pairwise distinct (\textit{someEq}), respectively.

As indicated by the subscript, our semantics \eval{.} makes use of an environment $E$,
which captures some relationships and properties of the RDD specification. We define
three functions over the environment:

\begin{tabbing}
xxx \= xxxxxxxxxxxxxxxxxxxxxx \= \kill
\>$E.C$ : IRI $\mapsto$ $List<$IRI$>$\>return the list of subclasses of a class,\\
\>\>according to the \verb!SUBCLASS! keyword\\
\>$E.P$ : IRI $\mapsto$ $List<$IRI$>$\>return the list of subproperties of a property,\\
\>\>according to the \verb!SUBPROPERTY! keyword\\
\>$E.A$ : IRI $\mapsto$ $List<$IRI$>$\>return the list of all properties defined in\\
\>\>any position inside a specific class definition
\end{tabbing}

For instance, if we have an RDD with only two class specifications \verb!CLASS A SUBCLASS B, C { ...}!
and \verb!CLASS C SUBCLASS D { ... }!, it holds that $E.C(A) = \{ B, C \}$, $E.C(C) = \{ D \}$ and for all
other $x \in IRI: E.C(x) = \emptyset$. Note that the environment functions are not defined in a
transitive way; instead, transitive inheritance of constraints along property and subclass hierarchies
is encoded directly in the inference rules where needed. The extraction of the environment from a
specific RDD along the grammar is straightforward, so we omit the technical details.

\newpage
\subsection{Top-level Decomposition}
The top-level rule decomposes the \textsc{RDD} specification into its two components, namely a list of class sections and
a property section.
The evaluation of the two sections results in a set of constraints, each, and the semantics of the \textsc{RDD}
is defined as the union of these two constraint sets.

{\small
\[
\inferrule{cs_{ccs} := $\eval{\textit{ccs}}$ \\ cs_{pcs} := $\eval{\textit{pcs}}$}
          {$\eval{(\textit{ccs}: ClassConstraintSec, \textit{pcs}: PropConstraintSec) : RDD}$\ \vdash cs_{ccs} \cup cs_{pcs}}
\]
}

\subsection{PropConstraintSec Decomposition}
\label{app:pcd}
The semantics of a \textsc{PropConstraintSec} is defined as the union over the individual constraints
contained in the section; in addition, the inference rule delegates to rule \textsc{OWA}$_P$ (defined subsequently),
which may set up a completeness constraint, according to the configuration. We write $p_i$.prop
to denote the property referred in the \textsc{PropConstraint} $p_i$ and indicate the use of the
\verb!OWA! keyword by the boolean variable \textit{isOWA} (where~\textit{isOWA}$=$\textbf{false} indicates
the use of keyword \verb!CWA!).

{\small
\[
\inferrule{ cs_{wa} := $\eval{([$p_1$.prop, \dots, $p_n$.prop], \textit{isOWA}) : OWA$_P$}$ \\ cs_{p_1} := $\eval{$p_1$}$ \\ \dots \\ cs_{p_n} := $\eval{$p_n$}$}
          {$\eval{([$p_1$ : PropConstraint, \dots, $p_n$ : PropConstraint], \textit{isOWA} : boolean) : PropConstraintSec}$\ \\ \vdash cs_{wa} \cup cs_{p_1} \cup \dots \cup cs_{p_n}}
\]
}

\subsubsection{Unqualified OWA$_P$} The \textsc{OWA}$_P$ rule splits up into two cases: if (i) the \verb!OWA!
keyword is used, (cf.~first rule, where we inject \textit{isOWA}=\textbf{true}), no constraint is enforced; (ii)
otherwise, according to the second rule, we set up a constraint enforcing that the list of properties is complete,
i.e.~no other properties than those listed occur in the underlying RDF data.
 
{\small
\[
\inferrule{ }
          {$\eval{([$p_1$ : IRI, \dots, $p_n$ : IRI], \textbf{true}) : OWA$_P$}$\ \vdash \emptyset}
\]
}

{\small
\[
\inferrule{ }
          {$\eval{([$p_1$ : IRI, \dots, $p_n$ : IRI], \textbf{false}) : OWA$_P$}$\ \vdash \\ \{ \forall s, p, o(T(s,p,o) \rightarrow p=p_1 \lor \dots \lor p=p_n) \}}
\]
}

\subsubsection{Unqualified PropConstraint} When evaluated in the context of a \textsc{PropConstraintSec},
the contained \textsc{PropConstraint}s are unqualified, i.e.~they are not restricted to properties being
used in the context of a specific class instance. According to the grammar, a \textsc{PropConstraint} is
composed out of a list of \textsc{Constraint}s (such as \textsc{MinConstraint}, \textsc{MaxConstraint},
\textsc{DomainConstraint}, etc.) and a property, possibly with range type restriction (such as \verb!IRI!,
\verb!BNode!, etc.).  Hence, the semantics of the rule is twofold. First, it induces a \textsc{RangeTypeConstraint},
which enforces the specified range type for the property; second, for every constraint in the list,
it induces an \textsc{UnqualifiedPropConstraint}. The rules for these two types will then be listed subsequently.

{\small
\[
\inferrule{cs_{rt} := $\eval{(\textit{p},\textit{rt}) : RangeTypeConstraint}$ \\ cs_{up} := \bigcup_{c \in cl}$\eval{(\textit{p},\textit{c}) : UnqualifiedPropConstraint}$}
          {$\eval{(\textit{cl} : List\<Constraint>, (\textit{p} : IRI, \textit{rt} : RangeType) : IRIWithRangeType) : PropConstraint}$\ 
\\ \vdash cs_{rt} \cup cs_{up}}
\]
}

\subsubsection{RangeTypeConstraint} 
The unqualified range type restriction enforces the range type of a property, according to one of
the keywords \textbf{IRI}, \textbf{BNODE}, \textbf{RESOURCE}, or \textbf{LITERAL} specified in the
RDD specification. Hence, we end up with five rules covering these cases, the first one for the case
where no range type restriction is present.

{\small
\[
\inferrule{ }
          {$\eval{(\textit{prop} : IRI, \textbf{null}) : RangeTypeConstraint}$\ \vdash \emptyset}
\]
}

{\small
\[
\inferrule{ }
          {$\eval{(\textit{prop} : IRI, \textbf{IRI}) : RangeTypeConstraint}$\ \vdash \{ \forall s, o (T(s,prop,o) \rightarrow IRI(o)) \}}
\]
}

{\small
\[
\inferrule{ }
          {$\eval{(\textit{prop} : IRI, \textbf{BNODE}) : RangeTypeConstraint}$\ \vdash\\ \{ \forall s, o (T(s,prop,o) \rightarrow BNode(o)) \}}
\]
}

{\small
\[
\inferrule{ }
          {$\eval{(\textit{prop} : IRI, \textbf{RESOURCE}) : RangeTypeConstraint}$\ \vdash\\ \{ \forall s, o (T(s,prop,o) \rightarrow Resource(o)) \}}
\]
}

{\small
\[
\inferrule{ }
          {$\eval{(\textit{prop} : IRI, \textbf{LITERAL}) : RangeTypeConstraint}$\ \vdash\\ \{ \forall s, o (T(s,prop,o) \rightarrow Literal(o)) \}}
\]
}

\subsubsection{UnqualifiedPropConstraints} 
Defining the core part of the semantics, the specific rules implementing the \textsc{PropConstraint}s
generate the individual FOL constraints; we define one rule per constraint type.

\medskip
\medskip
\noindent
\textit{Unqualified MinConstraint} \\
The inference rule induces a constraint enforcing that a given property $p$ is present at least $n$ times for every resource.

{\small
\[
\inferrule{ }
          {$\eval{($p$ : IRI, ($n$ : Integer) : MinConstraint)}$\ \vdash \\ \{ \forall s (Resource(s) \rightarrow \exists o_1, \dots o_n (T(s,p,o_1) \land \dots \land T(s,p,o_n) \land \textit{allDist}(o_1,\dots,o_n))) \}}
\]
}

\medskip
\noindent
\textit{Unqualified MaxConstraint} \\
The inference rule induces a constraint enforcing that a given property $p$ is present at most $n$ times for every resource.

{\small
\[
\inferrule{ }
          {$\eval{($p$ : IRI, ($n$ : Integer) : MaxConstraint)}$\ \vdash \\ \{ \forall s, o_1, \dots, o_{n+1} (T(s,p,o_1) \land \dots \land T(s,p,o_{n+1}) \rightarrow \textit{someEq}(o_1,\dots,o_{n+1})) \}}
\]
}

\medskip
\noindent
\textit{Unqualified DomainConstraint} \\
The inference rule induces a constraint enforcing that all resources with outgoing edge $p$ have some given type $D$.

{\small
\[
\inferrule{ }
          {$\eval{($p$ : IRI, ($D$ : IRI) : DomainConstraint)}$\ \vdash \{ \forall s, o (T(s,p,o) \rightarrow T(s,\textbf{rdf:type},D)) \}}
\]
}

\medskip
\noindent
\textit{Unqualified RangeConstraint} \\
The inference rule induces a constraint enforcing that all resources with incoming edge $p$ have some given type $R$.

{\small
\[
\inferrule{ }
          {$\eval{($p$ : IRI, ($R$ : IRI) : RangeConstraint)}$\ \vdash \{ \forall s, o (T(s,p,o) \rightarrow T(o,\textbf{rdf:type},R)) \}}
\]
}

\medskip
\noindent
\textit{Unqualified PathConstraint} \\
The inference rule induces a constraint enforcing that, for every value referred to by property $p$, there exists a path along edges $q_1, \dots q_n$ pointing to the same value.

{\small
\[
\inferrule{ }
          {$\eval{($p$ : IRI, ([$q_1$,\dots,$q_n$] : List\<IRI>) : PathConstraint)}$\ \vdash \\ \{ \forall s, o (T(s,p,o) \rightarrow \exists o_1, \dots, o_{n-1} (T(s,q_1,o_1) \land T(o_1,q_2,o_2) \land \dots \land T(o_{n-1},q_n,o))) \}}
\]
}

\medskip
\noindent
\textit{Unqualified SubPropertyConstraint} \\
The inference rule induces, for every subproperty $p_s$ of $p$ and every triple $(s,p_s,o)$, a constraint
enforcing that triple $(s,p,o)$ is in the database. Additionally, it implements recursive generation of \textsc{SubPropertyConstraint}s along the subproperty hierarchy $E.P$ from the environment.

{\small
\[
\inferrule{cs_{sp} := \bigcup_{p_s \in subProps} \{ \forall s, o (T(s,p_s,o) \rightarrow T(s,p,o))\} \\ cs_{rec} := \bigcup_{p_s \in \textit{subProps}}$\eval{($p$, ($E.P$($p_s$)) : SubPropertyConstraint)}$}
          {$\eval{($p$ : IRI, (\textit{subProps} : List\<IRI>) : SubPropertyConstraint)}$\ \vdash cs_{sp} \cup cs_{rec}}
\]
}

\medskip
\noindent
\textit{Unqualified PartialityConstraint} \\
The inference rule induces a constraint enforcing partiality of the property (i.e., it is either not present or single-valued, for every resource).

{\small
\[
\inferrule{cs_{part} := $\eval{(p,(\textbf{1}) : MaxConstraint)}$  }
          {$\eval{($p$ : IRI, () : PartialityConstraint)}$\ \vdash cs_{part}}
\]
}

\medskip
\noindent
\textit{Unqualified TotalityConstraint} \\
The inference rule induces a constraint enforcing totality of the property (i.e., it is always present exactly once, for every resource).

{\small
\[
\inferrule{cs_{total} := $\eval{($p$,(\textbf{1}) : MinConstraint)}$\ \cup\ $\eval{($p$,(\textbf{1}) : MaxConstraint)}$ }
          {$\eval{($p$ : IRI, () : TotalityConstraint)}$\ \vdash cs_{total}}
\]
}

%%%%%%%%%%%%%%%%%%%%%%%%%%%%%%%%%%%%%%%%%%%%%%%%%%%%%%%%%
%%%%%%%%%%%%%%%%%%%%%%%%%%%%%%%%%%%%%%%%%%%%%%%%%%%%%%%%%
%%%%%%%%%%%%%%%%%%%%%%%%%%%%%%%%%%%%%%%%%%%%%%%%%%%%%%%%%
\subsection{ClassConstraintSec Decomposition}
Like the \textsc{PropConstraintSec}, the \textsc{ClassConstraintSec} semantics is defined as the
union over the individual constraints contained in the section; analogously to \textsc{OWA}$_P$ in the 
\textsc{PropConstraintSec} rule, the call of OWA$_C$ induces a constraint according to the OWA/CWA
specification. We write $c_i.class$ for the class referred in \textsc{ClassConstraint} $c_i$.

{\small
\[
\inferrule{cs_{wa} := $\eval{([$c_1$.class, \dots, $c_n$.class], \textit{isOWA}) : OWA$_C$}$ \\ cs_{c_1} := $\eval{$c_1$}$ \\ \dots \\ cs_{c_n} := $\eval{$c_n$}$}
          {$\eval{([$c_1$ : ClassConstraint, \dots, $c_n$ : ClassConstraint], \textit{isOWA} : Boolean): ClassConstraintSec}$\ \\ \vdash cs_{wa} \cup cs_{c_1} \cup \dots \cup cs_{c_n}}
\]
}

\subsubsection{OWA$_C$}
If the \textit{isOWA} flag (second parameter) is set to \textbf{true}, no constraint is derived (first rule);
otherwise, we set up a constraint enforcing that the specified list of classes w.r.t.~instances in the data set.

{\small
\[
\inferrule{ }
          {$\eval{([$c_1$ : IRI, \dots, $c_n$ : IRI], \textbf{true}) : OWA$_C$}$\ \vdash \emptyset}
\]
}

{\small
\[
\inferrule{ }
          {$\eval{([$c_1$ : IRI, \dots, $c_n$ : IRI], \textbf{false}) : OWA$_C$}$\ \vdash \\ \{ \forall s, c(T(s,\textbf{rdf:type},c) \rightarrow c=c_1 \lor \dots \lor c=c_n) \}}
\]
}

\subsubsection{ClassConstraint}
The inference rule derives constraints for the individual components of
a class (keys and property constraints) and for the global configuration of the class
(i.e., subclass hierarchy, whether it is defined
as \verb!SINGLETON!, and whether it is defined as \verb!OWA!/\verb!CWA!). As usual, we indicate the presence of
the \verb!SINGLETON! and \verb!OWA! keywords by boolean variables \textit{isSingleton} and \textit{isOWA}.
We refer the interested reader to the main part of the paper for more details.

{\small
\[
\inferrule{cs_{singleton} := $\eval{(\textit{class}, \textit{isSingleton}) : Singleton}$ \\
               cs_{sc} := \bigcup_{c_{sc} \in \textit{subClasses}} $\eval{($c_{sc}$,\textbf{false},$E.C$($c_{sc}$),\textit{keys},\textit{cpcs},\textbf{true}) : ClassConstraint}$ \\
               cs_{key} := \bigcup_{key \in keys} $\eval{(\textit{class},\textit{key}) : ClassKey}$ \\ 
               cs_{qpcs} := \bigcup_{cpc \in cpcs} $\eval{(\textit{class},\textit{cpc}) : ClassPropConstraint}$ \\
               cs_{wa} := $\eval{(\textit{class\textit}, $E.A$(\textit{class}), \textit{isOWA}) : OWA$_P$}$}
          {\llbracket$(\textit{class} : IRI, \textit{subClasses} : List\<IRI>, \textit{isSingleton} : Boolean, \textit{keys} : List\<Key>,$ \\
             $\textit{qpcs} : List\<PropConstraint>, \textit{isOWA} : Boolean) : ClassConstraint$\rrbracket \vdash \\
             cs_{singleton} \cup cs_{sc} \cup cs_{key} \cup cs_{qpcs} \cup cs_{wa}}
\]
}

\subsubsection{Singleton} 
If the class is defined as \textbf{SINGLETON} (indicated by injecting \textbf{true} into the first rule),
a constraint is set up enforcing that the class has exactly one instance; otherwise, no constraint is derived.

{\small
\[
\inferrule{ }
          {$\eval{($C$ : IRI, \textbf{true}) : Singleton}$\ \vdash \{ \exists s( T(s,\textbf{rdf:type},C)) \}\ \cup \\ \{ \forall s_1, s_2(T(s_1,\textbf{rdf:type},C) \land T(s_2,\textbf{rdf:type},C) \rightarrow s_1=s_2 \}}
\]
}

{\small
\[
\inferrule{ }
          {$\eval{($C$ : IRI, \textbf{false}) : Singleton}$\ \vdash \emptyset}
\]
}

\subsubsection{Qualified OWA$_P$}
The rule is similar to its unqualified counterpart (except that the restriction covers only to instances of a given class).

{\small
\[
\inferrule{ }
          {$\eval{($C$ : IRI, [$p_1$,\dots,$p_n$] : List\<IRI>, \textbf{false}) : OWA$_P$}$\ \vdash \\ \{\forall s, p, o (T(s,\textbf{rdf:type},C) \land (s,p,o) \rightarrow p=p_1 \lor \dots \lor p=p_n)\}}
\]
}

\subsubsection{ClassPropConstraint}
The \textsc{ClassConstraint} decomposition rule delivers us \textsc{ClassPropConstraint}s, which are
covered by this rule; they are the qualified counterpart of the \textsc{PropConstraint}s discussed earlier, and hence working analogously.

{\small
\[
\inferrule{cs_{rt} := $\eval{($C$,($p$,\textit{rt})) : QualifiedRangeTypeConstraint}$ \\ cs_{qp} := \bigcup_{c\in cl}$\eval{($C$,($p$,$c$)) : QualifiedPropConstraint}$}
          {$\eval{($C$ : IRI, (\textit{cl} : List\<Constraint>, ($p$ : IRI, \textit{rt} : RangeType)) : PropConstraint) : ClassPropConstraint}$\\ \vdash cs_{rt} \cup cs_{qp}}
\]
}

\subsubsection{QualifiedRangeTypeConstraint} The rules are the qualified counterparts of the rules
for \textsc{RangeTypeConstraint}s presented earlier, and hence working analogously.

{\small
\[
\inferrule{ }
          {$\eval{($C$ : IRI, (\textit{prop} : IRI, \textbf{null})) : QualifiedRangeTypeConstraint}$\ \vdash \emptyset}
\]
}

{\small
\[
\inferrule{ }
          {$\eval{($C$ : IRI, (\textit{prop} : IRI, \textbf{IRI})) : QualifiedRangeTypeConstraint}$\ \vdash\\ \{ \forall s, o (T(s,\textbf{rdf:type},C) \land T(s,prop,o) \rightarrow IRI(o)) \}}
\]
}

{\small
\[
\inferrule{ }
          {$\eval{($C$ : IRI, (\textit{prop} : IRI, \textbf{BNODE})) : QualifiedRangeTypeConstraint}$\ \vdash\\ \{ \forall s, o (T(s,\textbf{rdf:type},C) \land T(s,prop,o) \rightarrow BNode(o)) \}}
\]
}

{\small
\[
\inferrule{ }
          {$\eval{($C$ : IRI, (\textit{prop} : IRI, \textbf{RESOURCE})) : QualifiedRangeTypeConstraint}$\ \vdash\\ \{ \forall s, o (T(s,\textbf{rdf:type},C) \land T(s,prop,o) \rightarrow Resource(o)) \}}
\]
}

{\small
\[
\inferrule{ }
          {$\eval{($C$ : IRI, (\textit{prop} : IRI, \textbf{LITERAL})) : QualifiedRangeTypeConstraint}$\ \vdash\\ \{ \forall s, o (T(s,\textbf{rdf:type},C) \land T(s,prop,o) \rightarrow Literal(o)) \}}
\]
}

\subsubsection{QualifiedPropConstraint} The rules are similar in spirit to those for the \textsc{PropConstraint} discussed earlier, and hence defined analogously.

\medskip
\medskip
\noindent
\textit{Qualified MinConstraint}
{\small
\[
\inferrule{ }
          {$\eval{($C$ : IRI, ($p$ : IRI, ($n$ : Integer) : MinConstraint))}$\ \vdash \\ \{ \forall s (T(s,\textbf{rdf:type},C) \rightarrow \exists o_1, \dots o_n (T(s,p,o_1) \land \dots \land T(s,p,o_n) \land \textit{allDist}(o_1,\dots,o_n))) \}}
\]
}

\medskip
\noindent
\textit{Qualified MaxConstraint}
{\small
\[
\inferrule{ }
          {$\eval{($C$ : IRI, ($p$ : IRI, ($n$ : Integer) : MaxConstraint))}$\ \vdash \\  \{ \forall s, o_1, \dots, o_{n+1} (T(s,\textbf{rdf:type},C) \land T(s,p,o_1) \land \dots \land T(s,p,o_{n+1}) \rightarrow \textit{someEq}(o_1,\dots,o_{n+1})) \}}
\]
}

\medskip
\noindent
\textit{Qualified DomainConstraint}
{\small
\[
\inferrule{ }
          {$\eval{($C$ : IRI, ($p$ : IRI, ($D$ : IRI) : DomainConstraint))}$\ \vdash \\ \{ \forall s, o (T(s,\textbf{rdf:type},C) \land T(s,p,o) \rightarrow T(s,\textbf{rdf:type},D)) \}}
\]
}

\medskip
\noindent
\textit{Qualified RangeConstraint}
{\small
\[
\inferrule{ }
          {$\eval{($C$ : IRI, ($p$ : IRI, ($R$ : IRI) : RangeConstraint))}$\ \vdash \\ \{ \forall s, o (T(s,\textbf{rdf:type},C) \land T(s,p,o) \rightarrow T(o,\textbf{rdf:type},R)) \}}
\]
}

\medskip
\noindent
\textit{Qualified PathConstraint}
{\small
\[
\inferrule{ }
          {$\eval{($C$ : IRI, ($p$ : IRI,  ([$q_1$,\dots,$q_n$] : List\<IRI>) : PathConstraint))}$\ \vdash \\ \{ \forall s, o (T(s,\textbf{rdf:type},C) \land T(s,p,o) \rightarrow \\ \exists o_1, \dots, o_{n-1} (T(s,q_1,o_1) \land T(o_1,q_2,o_2) \land \dots \land T(o_{n-1},q_n,o))) \}}
\]
}

\medskip
\noindent
\textit{Qualified SubPropertyConstraint}
{\small
\[
\inferrule{cs_{sp} := \bigcup_{p_s \in subProps} \{ \forall s, o (T(s,\textbf{rdf:type},C) \land T(s,p_s,o) \rightarrow T(s,p,o))\} \\ cs_{rec} := \bigcup_{p_s \in \textit{subProps}}$\eval{($C$: IRI, ($p$, ($E.P$($p_s$)) : SubPropertyConstraint))}$}
          {$\eval{($C$ : IRI, $p$ : IRI, (\textit{subProps} : List\<IRI>) : SubPropertyConstraint))}$\ \vdash cs_{sp} \cup cs_{rec}}
\]
}

\medskip
\noindent
\textit{Qualified PartialityConstraint}
{\small
\[
\inferrule{cs_{part} := $\eval{($C$,($p$,(\textbf{1}) : MaxConstraint))}$  }
          {$\eval{($C$ : IRI, ($p$ : IRI, () : PartialityConstraint))}$\ \vdash cs_{part}}
\]
}

\medskip
\noindent
\textit{Qualified TotalityConstraint}
{\small
\[
\inferrule{cs_{total} := $\eval{($C$,($p$,(\textbf{1}) : MinConstraint))}$\ \cup\ $\eval{($C$,($p$,(\textbf{1}) : MaxConstraint))}$ }
          {$\eval{($C$ : IRI, ($p$ : IRI, () : TotalityConstraint))}$\ \vdash cs_{total}}
\]
}

\subsubsection{ClassKey}
An $n$-ary class key over $n$ properties (with $n\geq 1$) induces (i) a range type constraint for every property
(given that the properties specified in the key may have a range associated according to the grammar),
(ii) totality of all key properties, and (iii) one constraint enforcing that no two distinct resources that are
member of the class coincide in all the values referred by the key properties. Note that, for now, the
RDD language does not support keys with multi-valued properties (which, as we argue, should
be a rare special case in practice).

{\small
\[
\inferrule{cs_{rt_1} := $\eval{($C$,($p_1$,\textit{rt$_1$})) : QualifiedRangeTypeConstraint}$ \\ \ \ \ \ \ \ \ \ \ \ \ \ \ \ \ \ \ \ \ \dots \ \ \ \ \ \ \ \ \ \ \ \ \ \ \ \ \ \ \ \ \\ cs_{rt_n} := $\eval{($C$,($p_n$,\textit{rt$_n$})) : QualifiedRangeTypeConstraint}$ \\ cs_{total_1} := $\eval{($C$,($p_1$,() : TotalityConstraint))}$\ \\ \ \ \ \ \ \ \ \ \ \ \ \ \ \ \ \ \ \ \ \ \ \ \ \dots \ \ \ \ \ \ \ \ \ \ \ \ \ \ \ \ \ \ \ \ \ \ \  \\ cs_{total_n} := $\eval{($C$,($p_n$,() : TotalityConstraint))}$}
          {$\eval{($C$ : IRI, [($p_1$,$rt_1$),\dots,($p_n$,$rt_n$)] : List\<IRIWithRangeType>) : ClassKey}$\ \vdash \\ cs_{rt_1} \cup \dots \cup cs_{rt_n} \cup cs_{total_1} \cup \dots \cup cs_{total_n} \cup \\ \{ \forall s_1, s_2, o_1, \dots, o_n(T(s_1,\textbf{rdf:type},C) \land T(s_1,p_1,o_1)\  \land \dots \land T(s_1,p_n,o_n)\ \land \ \ \ \ \ \ \ \ \ \ \ \ \ \\ \ \ \ \ \ \ \ \ \ \ \ \ \ \ \ \ \ \ \ \ \ \ \ \ \ \ T(s_2,\textbf{rdf:type},C) \land T(s_2,p_1,o_1) \land \dots \land T(s_2,p_n,o_n) \rightarrow s_1=s_2) \}}
\]
}
 \end{appendix}
\fi

\end{document}